\documentclass[a4paper,fleqn]{cas-dc}

\usepackage[numbers]{natbib}

\def\tsc#1{\csdef{#1}{\textsc{\lowercase{#1}}\xspace}}
\tsc{WGM}
\tsc{QE}
\tsc{EP}
\tsc{PMS}
\tsc{BEC}
\tsc{DE}

\begin{document}
\let\WriteBookmarks\relax
\def\floatpagepagefraction{1}
\def\textpagefraction{.001}

\shorttitle{Pore-scale Imaging and Multi-scale Modelling of Porous Building Materials}

\shortauthors{Menke, H.P. et~al.}

\title [mode = title]{Multi-scale flow, permeability, and heat transport in low-carbon and traditional building materials}                      
\tnotemark[1]
\tnotemark[2]
\tnotemark[3]
\tnotetext[1]{This document is the results of research funded by the EPSRC additive manufacturing grant.}
\tnotetext[2]{H.P.M.(First Author) contributed equally to this work with J.M. (Last Author)}
\tnotemark[3]{H.P.M, J.M., and G.M.M. conceptualised this study, K.M.H and H.P.M chose the samples, H.P.M. and K.S. did the imaging and analysis, J.M. and H.M. built the software and ran the models, All authors contributed to writing the paper.  }
%

\author[1]{Hannah P. Menke}[type=editor,
                        auid=000,bioid=1,
                        orcid=0000-0002-1445-6354]

\cormark[1]


\ead{h.menke@hw.ac.uk}



\affiliation[1]{organization={Institute of GeoEnergy Engineering},
    addressline={Heriot-Watt University}, 
    city={Edinburgh},
    postcode={EH14 4AS}, 
    country={United Kingdom}}



\author[3,4]{Katherine M. Hood}

\author[1]{Kamaljit Singh}

\author[2]{Gabriela M. Medero}

\author[1]{Julien Maes}


\affiliation[2]{organization={Institute for Sustainable Built Environment},
    addressline={Heriot-Watt University}, 
    city={Edinburgh},
    postcode={EH14 4AS}, 
    country={United Kingdom}}


\affiliation[3]{organization={Narro Associates},
    addressline={Orchard Brae House, 30 Queensferry Rd}, 
    city={Edinburgh},
    postcode={EH4 2HS}, 
    country={United Kingdom}}

\affiliation[4]{organization={Society for the Protection of Ancient Buildings},
    addressline={37 Spital Square}, 
    city={London},
    postcode={E1 6DY}, 
    country={United Kingdom}}
    
\cortext[cor1]{Corresponding author}


\begin{abstract}
Permeability and heat transport through building materials ultimately dictates their insulatory performance over a buildings service lifetime. However, characterisation of building materials is challenging because porous building materials are heterogeneous and their macroscopic physical properties (e.g. permeability, thermal, and mechanical behaviour) depend on their micro scale characteristics, i.e. the local distribution, material's fabric, and features of the solid components and the connectivity of the spaces between them. Large-scale testing can measure these macro-scale properties, but often does not give insight into the underlying  micro structural properties that ultimately leads to optimisation. Thus, a knowledge of the 3D structure is therefore required to assist in the development and implementation process for new materials. Experiments combining X-ray microtomography with numerical modelling are an accepted method of studying pore scale processes and have been used extensively in the oil and gas industry to study highly complex reservoir rocks. However, despite the obvious similarities in structure and application, these techniques have not yet been widely adopted by the building and construction industry.

An experimental investigation was performed on the pore structure of several building materials, including conventional, historic, and innovative, using X-ray tomography and direct numerical simulation. Six samples were imaged at between a 4 and 15 $\mu$m resolution inside a micro-CT scanner. The porosity and connectivity were extracted with the grain, throat, and pore size distributions using image analysis. The permeability, velocity, and thermal conductivity were then investigated using GeoChemFoam, our highly-versatile and open source numerical solver. It was found that each material had a unique, heterogeneous and sometimes multi-scale structure that had a large impact on the permeability and thermal conductivity. Furthermore, it was found that the method of including sub resolution porosity directly effected these bulk property calculations for both parameters, especially in the materials with high structural heterogeneity. This is the first multi-scale study of structure, flow and heat transport on building materials and this workflow could easily be adapted to understand and improve designs in other industries that use porous materials such as fuel cells and batteries technology, lightweight materials and insulation, and semiconductors.

\end{abstract}


\begin{highlights}
\item First multi-scale study of both low carbon and traditional building materials.
\item Micro-CT imaging and image analysis of building materials.
\item Darcy-Brinkman-Stokes permeability, flow, and  heat transport.
\item Conjugate thermal conductivity.
\end{highlights}

\maketitle
\section{Introduction}

The Intergovernmental Panel on Climate Change (IPCC) has stated that reducing carbon emissions from the built environment, which account for almost 40 percent of the total carbon emissions globally,  will make a meaningful impact on global climate change\cite{metz2005ipcc,iea2019global}. The global building stock continues to rise annually and strategies are being developed to mitigate the associated increases in emissions and achieve net zero carbon, largely by reducing building emissions \cite{twinn2019net}. Around 25 percent of embedded carbon in construction is generated in the building material production, transport, and construction stages \cite{gan2017developing}. One method of reducing carbon during the construction process is through the use of recycled or circularly produced materials that use less carbon and energy and are resource efficient to produce than traditional building materials \cite{etxeberria2007recycled, robayo2017alkali,medero2020construction}.  However, as these building materials are assimilated into the construction industry, their properties must be measured both for optimisation during the material development process and for accurate estimation during building design and construction. 

Characterisation of building materials is challenging for several reasons. Porous building materials are heterogeneous materials and their macroscopic physical properties (e.g. permeability, thermal, and mechanical) depend on their micro scale characteristics, i.e. the local distribution and features of the solid components and the connectivity of the spaces between them. For example, the porosity and permeability of building materials are ultimately related to the underlying structure and arrangement of the grains (solid particles), pores, and binding agent. Thus, two concretes made of the same materials, but with different sizes of aggregate can have vastly different thermal, transport, and mechanical properties \cite{dos2003effect}. Porous building materials also often contain a binding agent such as cement, which is itself porous, and has different material properties than the aggregate. The average pore throat size of the cement is often also on a length-scale orders of magnitude lower than the pores between the solid grains which makes it difficult to estimate macro-scale properties of the building material prior to testing. Large-scale testing can measure these macro-scale properties, but often does not give insight into the underlying micro structural properties that ultimately leads to optimisation and thus a knowledge of the 3D structure and particle arrangement is  required to assist in the development, optimisation, and implementation process for new low carbon building materials.  

Experiments combining X-ray microtomography ($\mu$-CT) with numerical modelling are now an accepted method of studying pore scale processes and have been used extensively in the oil and gas and carbon storage industries to study highly complex reservoir rocks \cite{blunt2013pore,noiriel2015resolving}. Pore-scale imaging experiments coupled with simulation are an increasingly important tool used in industry prediction of geological and petrophysical properties including porosity and connectivity \cite{menke2019using}, mineralogical heterogeneity \cite{lai2015pore}, and relative permeability \cite{2017-reynolds,armstrong2016beyond}, and thermal conductivity \cite{maes2021geochemfoam,liu2016pore}. In addition, new numerical techniques such as Darcy-Brinkman-Stokes (DBS) have been implemented to tackle the multi-scale nature of rocks, where processes are modelled explicitly in the pores using Stokes, and implicitly as an averaged volume in the microporous regions \cite{menke2021upscaling,faris2020investigation, soulaine2016impact}. However, despite the obvious similarities in structure and application, these techniques have not yet been widely adopted by the building materials and construction industry. 

There have been several recent studies that have combined imaging and modelling  experimental and modelling workflows to study materials used in the building industry.  Bentz et al. \cite{bentz2000microstructure} showed the applicability of pore scale analysis to porous building materials by imaging a lime clinker brick and a hard-burned clay brick using ($\mu$-CT) and confirmed good agreement with the Katz-Thompson relationship of permeability, diffusivity, and pore-size. However, their imaging resolutions were insufficient to accurately compute these properties with numerical simulations. Nunes et al. \cite{nunes2017influence} used nuclear magnetic resonance imaging to investigate the influence of pore structure on moisture transport in lime-plaster-brick systems and found that linseed oil application hindered the drying of the brick. Chung et al.\cite{chung2016evaluation} evaluated the performance of glass beads on thermal conductivity in insulating concrete and confirmed the results using numerical modelling. Still, they used limited image analysis and did not measure the connectivity of the sample, instead relying on probability functions to describe the structural arrangement of beads and voids. In addition, while these studies have shown the potential for modern pore-scale analysis, none of them have incorporated multi-scale modeling techniques to improve accuracy and computational expense.

Several studies have also attempted to model heat transport in building materials. Bicer et al. \cite{bicer2023modelling} and Guo et al. \cite{guo2011thermal} built purely mathematical models that did not include any simulations or pore scale information.  Zhang et al. \cite{zhang2015mesoscale} built a theoreticial mesoscale model for thermal conductivity of concrete that was based on bulk properties and did not consider multi-scale heterogeneity. Zhu et al. \cite{zhu2023multiscale} built a multi-scale theoretical model for concrete and bench marked the results against simulations and experiments. However, the simulations were only 2D and did not explicitly model pores (only pores completely filled with aggregate) meaning that the heat conductivity along connected pathways of grains or empty pores could not be accounted for explicitly. Panerai et al. \cite{panerai2017micro} investigated thermal conductivity in a range of fibrous structures using microCT imaging and 3D modeling, but did not investigate the effect of the model used for effective thermal conductivity of the microporous fibres. Xiao et al \cite{xiao2023influence} used numerical modelling to investigate heat conduction of random porous fibres but the simulations were limited to 2D. Shen et al. \cite{shen2023insights} studied the thermal conductivity of various hybrid steel-fine polypropylene and fiber-reinforced concrete with theoretical models, experiments, and 2D simulations, finding that the touching solid components acted as a thermal bridge. Yet, 2D numerical models limited the complexity of these investigations. Yang et al., \cite{yang2019thermal} did numerical investigations into the effect of fractures on thermal conductivity. However, their fractures were simple and straight and their digital rock was not derived from a real 3D sample. Siegert et al \cite{siegert2021validation} were the first to point out the importance of understanding harmonic vs arithmetic averaging on heat transfer in porous media structures, showing vast difference in effective properties depending on which method was used. Nevertheless, their investigations were limited to meshing of simple, relatively homogeneous samples that did not contain multi-scale porosity. However, as of yet, no one has used multi-scale modelling to compare the permeability and thermal conductivity of different building materials with varying structural complexity and materiel components. 

The objective of this study is to showcase the potential of combined pore-scale imaging and multi-scale modelling in use-cases relevant to the building industry. (1) First, six samples of building materials are imaged inside a micro-CT scanner at a 4-15 micron resolution. (2) The images are then analysed for pore, throat, and grain size distributions, and the connectivity of the pore-space is assessed both with and without including the microporosity. (3) Permeability is then calculated on the images using the Darcy-Brinkman-Stokes method, both included and excluding the microporous phase. (4) Finally, the thermal conductivity inside the materials is estimated. As the microporosity is unresolved in our micro-CT images, its structure is unknown and its thermal conductivity needs to be modelled by correlations. Here, both harmonic and arithmetic averaging of the conductivity of the fluid and solid inside the microporous phase is used and the results compared. 

\section{Sample Selection}

Seven samples were selected for this study, five traditional building materials from various time periods, and a low carbon brick (KBriq) made from recycled construction and demolition waste and a proprietary binder by Kenoteq \cite{grose2022grown}. Five traditional materials were chosen: a fired clay brick from a 1930's building, a brick sample (approximately 1900's) drilled from the Mackintosh Building at the Glasgow School of Art (GSA), modern aerated concrete, a wooden beam from a 1930's building, and Bentheimer sandstone. These materials have a range of flow and heat transport properties with heterogeneous pore structures that make them ideal benchmark cases for modern multi-scale numerical solvers. 

\section{Imaging and Analysis Methods}

Cylindrical cores with a diameter of 9-mm were drilled from each of the samples, except the aerated concrete where a larger piece was extracted to ensure a representative volume due to the large distribution in grain sizes, and Bentheimer for which a 4 $\mu$m resolution image of which was downloaded from https://www.digitalrocksportal.org/. The samples were imaged in three dimensions inside an EasyTom ($\mu$-CT) scanner from Rx Solutions at 100keV and 10W at between a 6 and 15 $\mu$m resolution. 

\begin{figure}[ht]
\begin{center}
\includegraphics[width=0.48\textwidth]{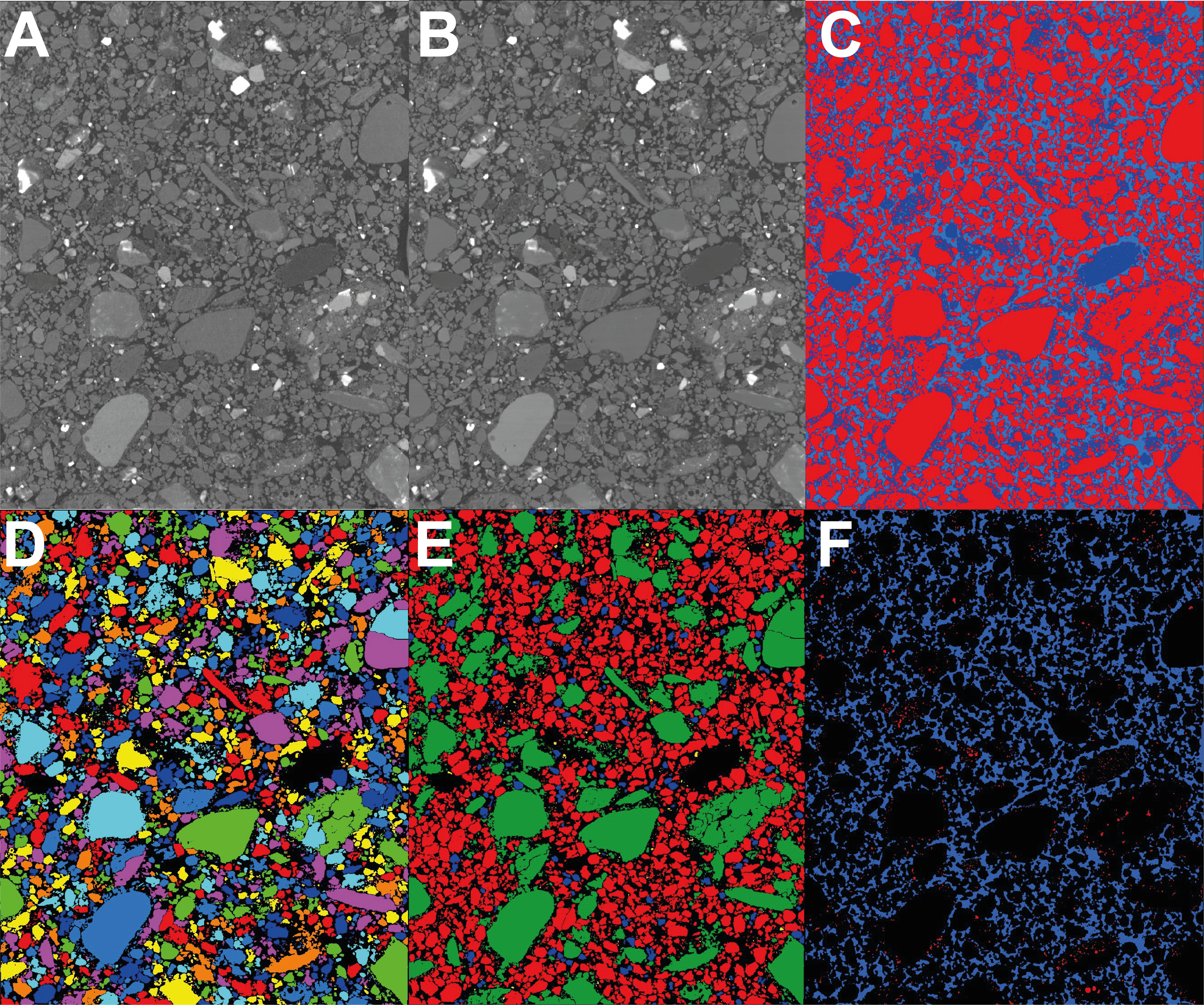}
\caption{ The raw image of KBriq (A), filtered image (B), segmentation (C) with the primary porosity rendered in light blue, microporosity in dark blue and solid grains in red shown for the KBriq sample. Grains are separated and randomly colored (D), and colored by size (E) with small grains in yellow (0-100$\mu$m equivalent diameter), medium sized grains in blue (100-200$\mu$m), large grains in red (200-500$\mu$m), and extra large grains in green (500+$\mu$m). Connected (blue) and unconnected (red) pores (A).   \label{fig:ImageAnalysis}}
\end{center}
\end{figure}

The raw projections were then reconstructed (Fig. \ref{fig:ImageAnalysis}A) and post-processed using the image processing modules in Avizo 2020.3.1 (www.thermofischer.com). The high-resolution micron-scale images were filtered using the non-local means filter \cite{buades2011non} (Fig. \ref{fig:ImageAnalysis}B) and segmented using the watershed segmentation algorithm \cite{beucher1993morphological} into three phases (pore, microporosity, and solid grain) (Fig. \ref{fig:ImageAnalysis}C). The pores, grains, and throats were then each separated into individual components using the separate objects module that uses the watershed technique on the Euclidean distance map of the phase (Fig. \ref{fig:ImageAnalysis}D) and independently analysed for size distributions (Fig. \ref{fig:ImageAnalysis}E). The pore space of each image was then analysed for connectivity in the Z direction and both connected and unconnected pores were identified (Fig. \ref{fig:ImageAnalysis}F). This connectivity analysis was done twice for each image, once for only the macropores, and again for the macropores and micropores together, giving an estimation of the connectivity of flow pathways both including and excluding the sub resolution porosity. The image analysis results are summarised in Table \ref{Table:ImageProperties}. This connectivity analysis is including the wooden beam where the cell walls themselves were assumed to be microporous \cite{stamm2002density} and thus no solid phase was identified in any segmentation. 

The average porosity of the microporosity was estimated by using the segmented microporosity as a mask for the grey-scale image. The histogram was then plotted for the microporous phase only. The histogram values were then normalised using the grey scale values used in watershed segmentation where the maximum grey-scale value of the macroporous phase was assigned to a porosity of 1 and the minimum grey-scale value of the solid phase was assigned to be a porosity of 0. A linear relationship between grey-scale value and porosity was assumed for the microporous phase and the mean value of the porosity histogram was used in the DBS numerical simulations as the porosity of the microporous phase.  

\section{Numerical Methods}

For each image, a $400^{3}$ voxel sub-volume was extracted from each image and used in all numerical simulations. Permeability was calculated using the DBS approach, both including and excluding the microporous phase. Steady state heat transfer (thermal conductivity) was then solved with a constant value for the heat conductivity of the solid $\kappa_s=1.4$ $kW/m/K$ and heat conductivity of the fluid $\kappa_f$ ranging from 0.0006 to 666 $kW/m/K$. All simulations used modules in GeoChemFoam 5.0 \cite{Maes2021,Maes2020b,Maes2021b}, our numerical toolbox based on OpenFOAM\textsuperscript{\textregistered} \cite{OpenFOAM2016}.The full code can be downloaded free of charge from \href{http://github.com/geochemfoam}{https://github.com/GeoChemFoam}. 

\subsection{Calculation of Permeability}

Flow in the images was solved using the \textit{simpleDBSfoam} module using the DBS approach  \cite{soulaine2016impact} in which one equation is used to model the flow within the fully resolved macropores (i.e. voxel porosity equal to 1.0) and the micropores (i.e. voxel porosity lower than 1.0).

\begin{equation}\label{Eq:momentumDBS}
 \mu\nabla^2\mathbf{u}-\nabla\mathbf{p}-\mathbf{\mu}k^{-1}\cdot\mathbf{u} = 0,
\end{equation}

\begin{equation}\label{Eq:velocity}
\nabla\cdot\mathbf{u} = 0,
\end{equation}
where $\mathbf{u}$ [m.s$^{-1}$] is the fluid velocity, $\mathbf{p}$ [Pa] is the pressure gradient, $\mu$ [kg.m$^1$.s$^{-1}$] is the fluid viscosity and $k$ [m$^2$] is the local permeability, i.e. the permeability of the computational cell. This coefficient is assigned to a large value of 10$^{13}$ m$^2$ in the pores and to a very small value of 10$^{-21}$ m$^2$ in the solid phase in order to obtain a no-flow no-slip condition at the fluid-solid interface.  

The  permeability of each microporous voxel is assigned using the Kozeny-Carmen relationship, which assumes that the microporosity consists of an even packing of equally-sized elliptical grains.  As there is no information about the size of the grains in the microporous phase \emph{a priori}, the grain size of this phase was taken from literature values of typical grain size distributions for each material.  The permeability of the microporous phase was estimated using the equation 
\begin{equation}\label{Eq:Kmicro}
K_{\mu} = \frac{h^{2}}{180}\frac{\phi_\mu^{3}}{(1-\phi_\mu)^{2}},
\end{equation}
where $\phi_\mu$ is the porosity of the microporous phase and $h$ is the average grain radius. Additional information on Darcy-Brinkman-Stokes modelling methods can be found in \cite{menke2021upscaling}.

To calculate the overall permeability $K$ of the sample, a pressure drop $\Delta P$ is applied between the left and the right boundaries, and the velocity field is calculated. The permeability is then obtained as
\begin{equation}
    K=-\frac{U_DL}{\mu \Delta P}
\end{equation}
where $L$ is the length of the sample and $U_D$ (m/s) is the Darcy velocity, defined as
\begin{eqnarray}
U_D=\frac{Q}{A},
\end{eqnarray}
where $A$ (m$^2$) is the cross-sectional area of the domain and $Q$ is the flow rate (m$^3$/s). 

\subsection{Calculation of effective heat conductivity}

Heat transport was solved using a simplified temperature equation,
\begin{equation}\label{Eq:Teff}
0 =  \nabla \cdot \overline{\kappa}\nabla T
\end{equation}
where $\overline{\kappa}$ is the local heat conductivity, i.e. the heat conductivity of the computational cell. This coefficient must include the contribution of the pores, micropores and solid present inside the computational cell. Inside the solid phase, $\overline{\kappa}=\kappa_s$, and inside the pores, $\overline{\kappa}=\kappa_f$. In the microporous phase, $\overline{\kappa}$  will depend on the heat conductivities of the fluid and solid, of the porosity $\phi_\mu$, and of the underlying structure of the pores inside the microporous phase. Calculating this coefficient accurately would require a high resolution image of this underlying structure. In the absence of such image, the local heat conductivity must be modelled. In this work, the results are compared using two different models:

\begin{equation}\label{Eq:Kmicroharm}
 \mathbf{\kappa}_{\mu(harmonic)}  = \frac{\mathbf{\kappa}_{f}\mathbf{\kappa}_{s}}{((1-\mathbf{\phi}_{\mu})\mathbf{\kappa}_{f}+\mathbf{\kappa}_{s}\mathbf{\phi}_{\mu})} 
\end{equation}

\begin{equation}\label{Eq:Kmicroerith}
 \mathbf{\kappa}_{\mu(arithmetic)}  = (1-\mathbf{\phi}_{\mu})\mathbf{\kappa}_{s}+\mathbf{\phi}_{\mu}\mathbf{\kappa}_{f}
\end{equation}

Additional information on the $heatTransportFoam$ solver numerical methods in GeoChemFoam can be found in \cite{maes2021geochemfoam}.

Upscaling heat transport requires calculating the effective heat conductivity coefficient  $\kappa_{eff}$  [$kW/m/K$] between the fluid and the solid, defined by \cite{siegert2021validation}. The effective heat conductivity regroups the contribution of fluid, solid and microporous phase into one single-field coefficient. To calculate it, a temperature drop $\Delta T$ is established between the left and right boundaries, the temperature field inside the domain is calculated and the effective heat conductivity coefficient is obtained as,
\begin{equation}
    \kappa_{eff}=\frac{Q_hL}{A\Delta T},
\end{equation}
where $Q_h$ is the heat flow rate, calculated as the integrate of the heat flux $\mathbf{J}=-\overline{\kappa}\nabla T$ across the cross secitonal area A.

\section{Results and Discussion}
\subsection{Image Analysis}
\begin{figure*}[h]
\begin{center}
\includegraphics[width=0.80\textwidth]{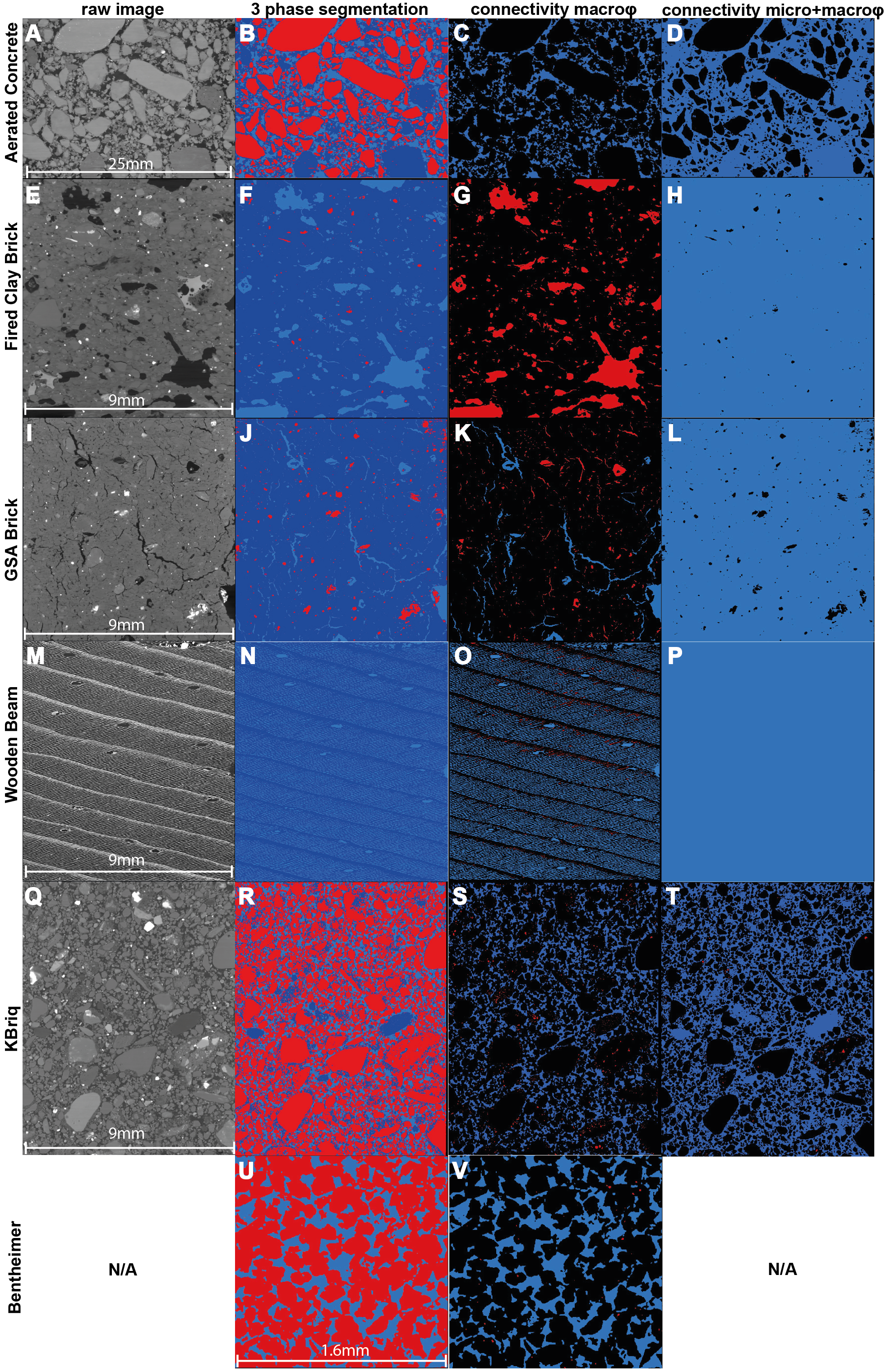}
\caption{ The raw (A, E, I, M, Q, U) , 3-phase segmentation (B,F,J,N,R,V,Y), connectivity of the macro porosity (C,G,K,O,S,W,Z), and the connectivity of the micro+macro porosity (D,H,L,P,T,X) of each sample. In the 3-phase segmentation, macro pores are light blue, micro pores are dark blue, and solid grains are red. In the connectivity of the macro porosity and macro+micro porosity, unconnected pores are red and connected pores are blue.  Bentheimer does not have microporosity and the image was obtained pre-segmented. \label{fig:ConnectivityAnalysis}}
\end{center}
\end{figure*}

The images were first segmented into three phases: macro pore, micro pore, and solid grain as show in Fig. \ref{fig:ConnectivityAnalysis}. The volume fraction of each phase is shown in Table \ref{Table:ImageProperties}. The connectivity of each phase of porosity was then assessed with just the macro porosity and then the connected macro+micro porosity. 

In this analysis the aerated concrete shows similar volume fractions to the KBriqs with 0.18-0.20 macro porosity, 0.18-0.38 micro porosity, and 0.42-0.64 solid grain. This high macroporosity volume fraction led to a high connectivity of macroporosity of 0.99 for all three samples. The GSA Brick and Fired Clay Brick each have large volume fractions of microporosity of 0.89 with little microporosity of 0.06-0.10 and little solid grain (0.01-0.05). The lack of macroporosity led to zero connectivity in the macroporosity in the Fired Clay Brick. However, in the GSA Brick, the porosity was present predominantly as long, thin fractures, leading to a relatively high connected volume fraction of the macroporosity of 0.3. In the wooden beam the porosity was present in laminated layers of pores with low porosity regions between. However, the connectivity was high regardless with a connected macroporosity of 0.97. Bentheimer did not contain microporosity but was almost entirely connected due to its strong homogeneity and large grains. 

\begin{figure*}[h]
\begin{center}
\includegraphics[width=1\textwidth]{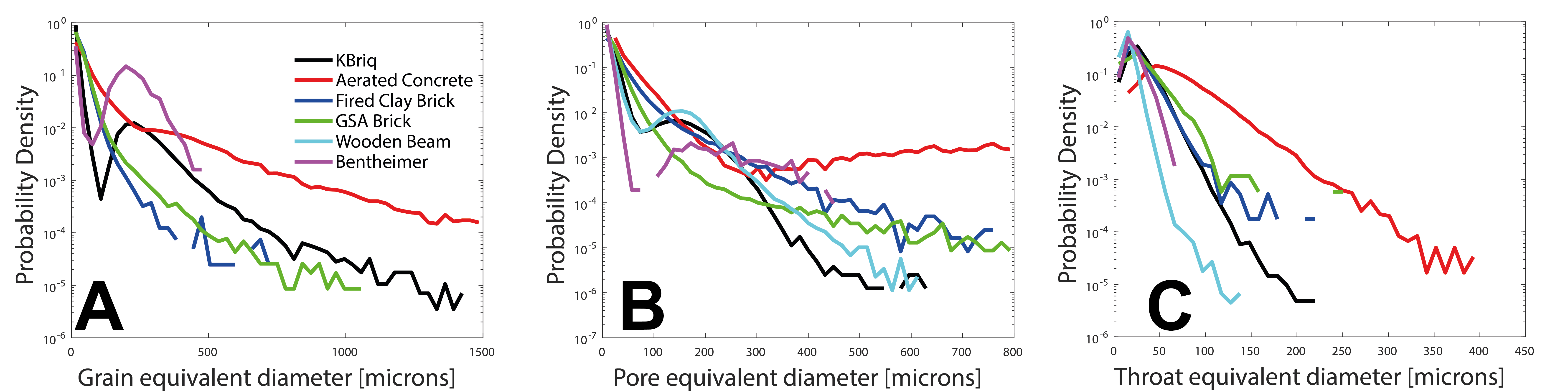}
\caption{ The grain size distributions (A), pore size distributions (B), and throat size distributions (C) of the samples. Note that the grain size distribution of the wooden beam is excluded because it does not contain grains.    \label{fig:PSD}}
\end{center}
\end{figure*}

The images were then analysed for pore, grain, and throat size distributions (Fig. \ref{fig:PSD}).  Aerated concrete had the largest grains with large numbers of grains even at the larger radii (Fig. \ref{fig:PSD}A). The grain size distribution of the KBriq are very similar to Aerated Concrete, with a peak in grain radius of around 250 microns. Bentheimer had the smallest variance in grain size with a peak at 250 microns. The Clay Fired and GSA bricks had a small number of isolated grains that were all under 500 microns. No grains were extracted from the wooden beam as all of its solid components were microporous. 

When comparing pore size (Fig. \ref{fig:PSD}B), Aerated Concrete had the largest pores with an even distribution stretching from 10-1500 microns. The other samples all had peaks between 100 and 200 microns, except the GSA brick which had a long tail indicative of the large cracks that can be seen throughout the sample. It is interesting to note that the Wooden Beam shows a peak much larger than would be expected for its small pores, which is due to the long pores stretching end to end of the sample and the calculation of pore equivalent diameter rather than maximal ball radius \cite{dong2009pore}.

The throat size distributions are shown in Fig. \ref{fig:PSD}C. Wooden beam and Bentheimer have the smallest pore throats with a peak around 20 microns and a very narrow distribution. Aerated concrete has a very wide distribution with a peak around 100 microns. The rest of the samples peak around 50 microns and have a small tail at larger throat sizes indicating some heterogeneity within the samples.

\begin{table*}[h]
\footnotesize
\centering
\newcommand{\wrap}[1]{\parbox{.15\linewidth}{\vspace{1.5mm}#1\vspace{1mm}}}
\begin{tabular}{c|c|c|c|c|c|c|c|c|c}
Sample & resolution & macro $\mathbf{\phi}$ &  micro $\mathbf{\phi}$ &  solid grain  & connected $\mathbf{\phi}$ & connected $\mathbf{\phi}$ & micro $\mathbf{\phi}$ & $h$ & $\mathbf{K_{\mu}}$ \\[0.005cm]
& $\mathbf{\mu}$m & vol Frac & vol Frac & vol Frac & macro & macro+micro & avg porosity & $\mathbf{\mu}$m & m$^{2}$ \\[0.05cm]
\hline 
& & \\[-0.2cm]
Aerated Concrete & 15 & 0.20 & 0.38 & 0.42 & 0.99 & 0.99 & 0.41 & 5 \cite{bentz1999effects} & 2.75 x $10^{-14}$ \\[0.05 cm]
Fired Clay Brick & 7 &  0.06 & 0.48 & 0.43 & 0.00 & 1.00 & 0.27 & 6.5 \cite{baspinar2010utilization} & 1.10 x $10^{-14}$ \\[0.05 cm]
GSA Brick & 7 & 0.06 & 0.89 & 0.05 & 0.33 & 1.00 & 0.34 & 6.5 \cite{baspinar2010utilization} & 2.12 x $10^{-14}$ \\[0.05 cm]
Wooden Beam & 7 & 0.34 & 0.66 & 0.0 & 0.97 & 1.0 & 0.72 & N/A & 2.70 x $10^{-20}$ \cite{petty1983permeability}** \\[0.05 cm]
KBriq & 8 & 0.18 & 0.18 & 0.64 & 0.955 & 0.996 & 0.35 & 5* & 1.41 x $10^{-14}$ \\[0.05 cm]
Bentheimer & 4 & 0.26 & 0 & 0.74 & 0.99 & N/A & N/A & N/A & N/A  \\[0.05 cm]
\end{tabular}
\caption{Porosity and properties of the samples measured using image analysis and estimated from literature values.*The KBriq samples were estimated to have a microporous grain size $h$ of 5 $\mathbf{\mu}$m due to its composition of construction and demolition inert waste. **The permeability of the microporous phase of wood is analogous to the permeability of a cell wall. \label{Table:ImageProperties}}
\end{table*}

\subsection{Numerical Simulation: Permeability with Darcy-Brinkman-Stokes}

\begin{figure*}
\begin{center}
\includegraphics[width=0.9\textwidth]{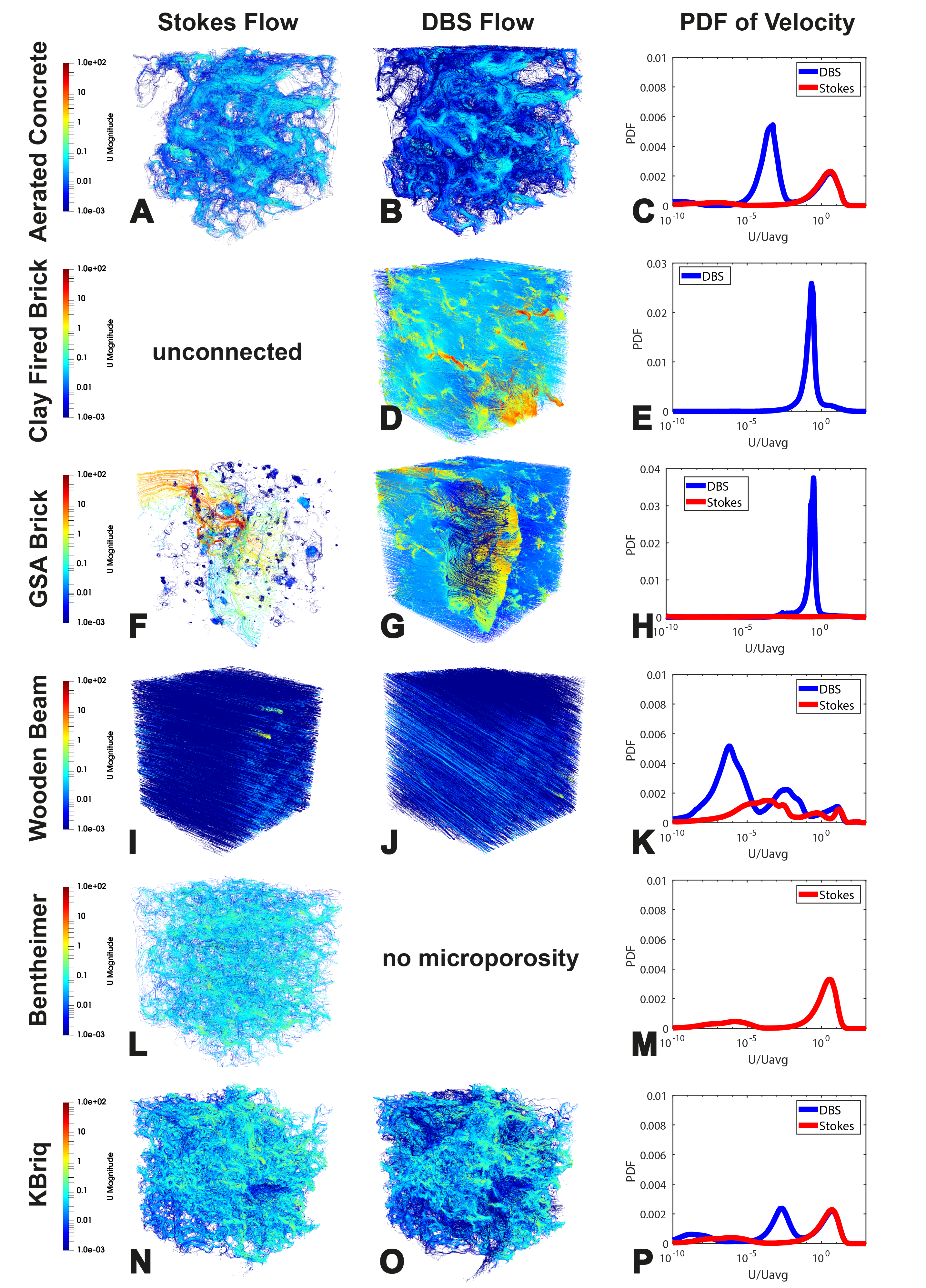}
\caption{ The velocity streamlines of each sample both excluding (Stokes) [A,F,I,L,N] and including (DBS) microporosity [B,D,G,J,O]. The probability density functions of $U/U_{avg}$ for each simulation are shown in column 3 [C,E,H,K,M,P].    \label{fig:VelocityStreamlines}}
\end{center}
\end{figure*}

The permeability $K$ was calculated on a $400^3$ voxel subvolume of the images both including and excluding the microporous phase using the DBS solver in GeoChemFoam. When excluding the microporosity both the solid and microporous phases were given a porosity of 0.0001 and a permeability of $10^{-20}$. When including the microporosity the average porosity of the microporosty was used along with the Kozeny-Carmen calculated $K_{micro}$, with the solid values remaining 0.0001 and $10^{-20}$ $m^2$. All calculated $\mathbf{\phi}$, $K$, $L$, and $U{_D}$  for each scenario are shown in Table \ref{Table:PermeabilityTransportPropertiesStokes}.  The streamlines for each scenario and the pds of velocity are shown in Fig.\ref{fig:VelocityStreamlines}. 

It is observed in Aerated Concrete (Fig.\ref{fig:VelocityStreamlines}. A, B, C) that the macro porous regions are highly connected with fast flow in many of the pores resulting in a moderate peak in the velocity PDF at high velocities. This peak is reduced and flattened with the inclusion of microporosity indicating that the binder is contributing to flow in some regions.  The permeability of Aerated concrete was the highest of all the samples at 5.4 x $10^{-11}$ $m^2$ , which is a reflection of the large pores and throat sizes, and increases less then 1 percent with the addition of microporosity.  The Clay Fired Brick (Fig.\ref{fig:VelocityStreamlines}. D, E) is not connected in the macro pores, however, when microporosity is included a low permeability of 5.9 x $10^{-14}$ $m^2$ is calculated and it is observed that flow is slow through the clay and then increases when an unconnected pore or grain is encountered. Nevertheless the difference between the minimum and maximum velocities is not more than one order of magnitude and thus only one peak is seen in the PDF. The GSA Brick (Fig.\ref{fig:VelocityStreamlines}.F, G, H) shows only a single connected path along the fracture with a permeability of 2 x $10^{-14}$ $m^2$.. The fraction of the sample volume containing high velocities is so low that the high velocities barely register on the PDF. When microporosity is included,  flow is fast in much of the sample as the micropores connect previously unconnected fracture networks and this is reflected in the permeability increase of 3.5x and the PDF with a strong peak at high velocities and a tail leading to lower velocities. 

The Wooden Beam (Fig.\ref{fig:VelocityStreamlines}. I, J, K) has pores that stretch end to end in the direction of flow, resulting in a flow field that is analogous to a bundle of tubes with a permeability of 5.9 x $10^{-13}$ $m^2$. For the Stokes case this results in several small peaks and troughs that depend on the various pore radii, with the few pores that are slightly bigger than the others becoming the preferential flow paths. When microporosity is added some of the smaller tubes become more connected and another peak is seen in the mid-range velocities and a small 1.2x increase in permeability.  Bentheimer  (Fig.\ref{fig:VelocityStreamlines}. L,M ) does not contain sub resolution porosity and is well connected resulting in only a single peak at high velocities and a permeability of 3.8 x $10^{-12}$ $m^2$. 

Finally, the KBriq  (Fig.\ref{fig:VelocityStreamlines}. Q,R,S ) a small peak at high velocities with Stokes flow and a similar permeability of 4  x $10^{-12}$ $m^2$. The peak in the PDF is then shifted to lower velocities and widened when microporosity is included due to connecting some stagnant pores that then contribute to over all flow, however, these connected change the permeability by less than 1 percent, which is attributed to the already high permeability and relatively homogeneous arrangement of pores and grains. 

It is therefore concluded that if the macropores are well connected, there are only minor effects on the flow field and permeability. However, if the macropores are poorly connected or disconnected then microporosity must be included in the flow calculations to solve for permeability.

\begin{table*}[h]
\footnotesize
\centering
\newcommand{\wrap}[1]{\parbox{.05\linewidth}{\vspace{1mm}#1\vspace{1mm}}}
\begin{tabular}{c|c|c|c|c|c|c|c|c}
Sample & $K_{Stokes}$ & $K_{DBS}$ & $\mathbf{\phi}_{Stokes}$ &  $\mathbf{\phi}_{DBS}$ & $L_{Stokes}$ & $L_{DBS}$ & $U_{D(Stokes)}$ & $U_{D(DBS)}$  \\[0.005cm]
& $[m^2]$ & $[m^2]$& [-] & [-] & [m] & [m] & $[m.s^{-1}]$ & $[m.s^{-1}]$ \\[0.05cm]
\hline 
& & \\[-0.2cm]
Aerated Concrete & 5.386 x $10^{-11}$ & 5.626 x $10^{-11}$ & 0.210 & 0.371 & 4.535 x $10^{-5}$ & 3.483 x $10^{-5}$ & 2.634 x $10^{-3}$ & 2.751 x $10^{-3}$\\[0.05 cm]
Fired Clay Brick & 0 (unconnected) & 5.92 x $10^{-14}$ & N/A & 0.365 & N/A & 1.319 x $10^{-6}$ & N/A & 1.215 x $10^{-1}$\\[0.05 cm]
GSA Brick & 2.022 x $10^{-14}$ & 7.13 x $10^{-14}$ & 0.039 & 0.359 & 2.033 x $10^{-6}$ & 1.26 x $10^{-6}$ & 1.20 x $10^{-2}$ & 4.230 x $10^{-2}$\\[0.05 cm]
Wooden Beam & 5.866 x $10^{-13}$ & 6.856 x $10^{-13}$ & 0.353 & 0.819  & 3.64 x $10^{-6}$& 2.589  x $10^{-6}$ & 2.103 x $10^{-3}$ & 2.460 x $10^{-3}$\\[0.05 cm]
KBriq & 4.033 x $10^{-12}$ &  4.125 x $10^{-12}$ & 0.179  & 0.245 & 1.33 x $10^{-5}$ &1.161 x $10^{-5}$ & 6.239x $10^{-3}$ & 6.382 x $10^{-3}$ \\[0.05 cm]
Bentheimer & 3.750 x $10^{-12}$ & N/A & 0.226  & N/A  & 1.150 x $10^{-5}$ & N/A  & 5.80 x $10^{-3}$ & N/A\\[0.05 cm]
\end{tabular}
\caption{Permeability and flow properties of the samples as calculated using GeoChemFoam's Darcy-Brinkman-Stokes Flow Solver (simpleDBSFoam) in 400$^{3}$ subvolumes. \label{Table:PermeabilityTransportPropertiesStokes}}
\end{table*}

\subsection{Numerical Simulations: Steady State Heat Transfer}

\begin{figure*}[b]
\begin{center}
\includegraphics[width=1\textwidth]{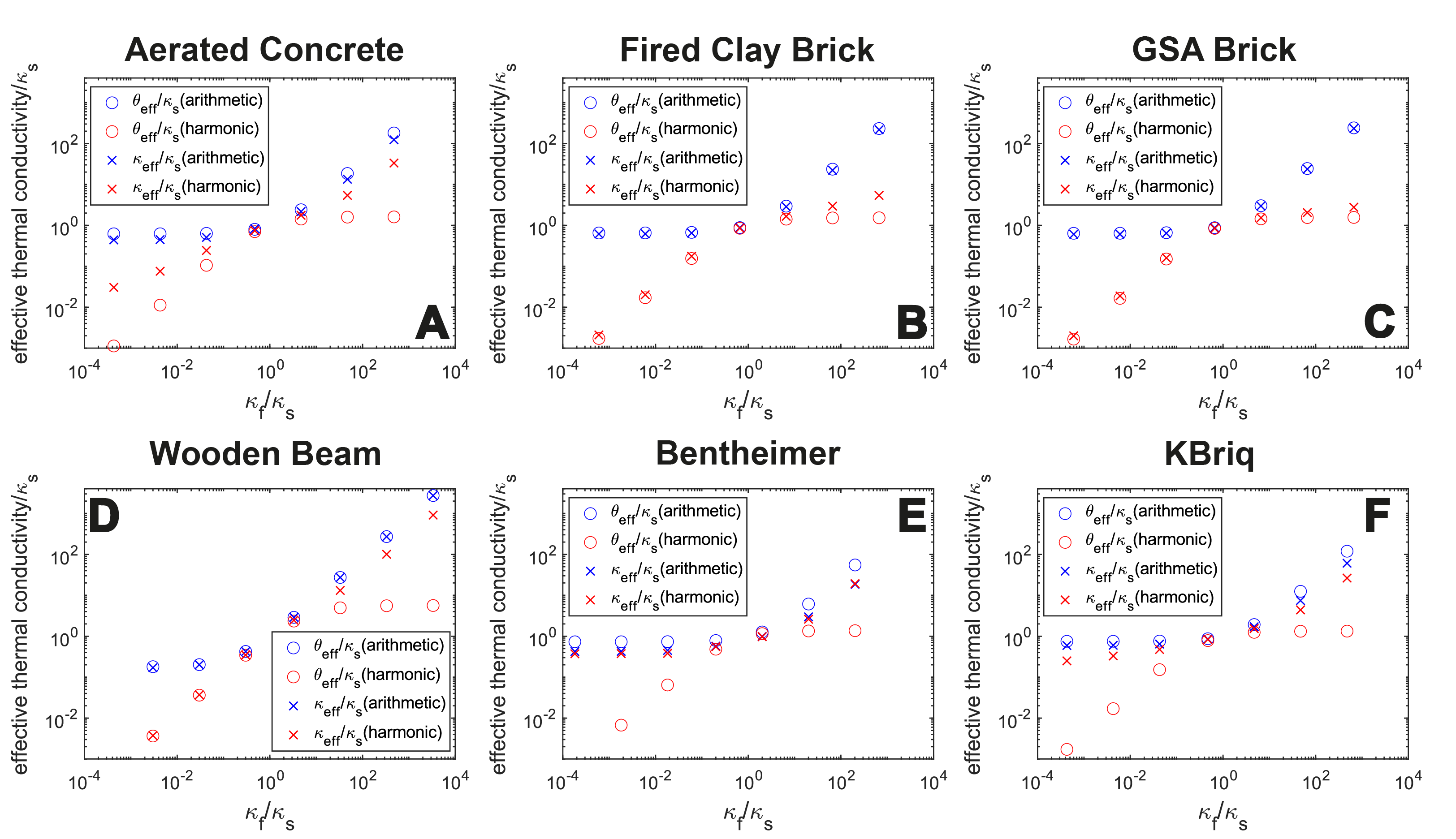}
\caption{ The $\mathbf{\kappa}_{eff}$ and $\mathbf{\theta}_{eff}$ normalised to $\mathbf{\kappa}_{s}$ for each of the building materials using both harmonic and arithmetic averaging of the components of the solid phase. \label{fig:Keff}}
\end{center}
\end{figure*}

\begin{figure*}[t]
\begin{center}
\includegraphics[width=1\textwidth]{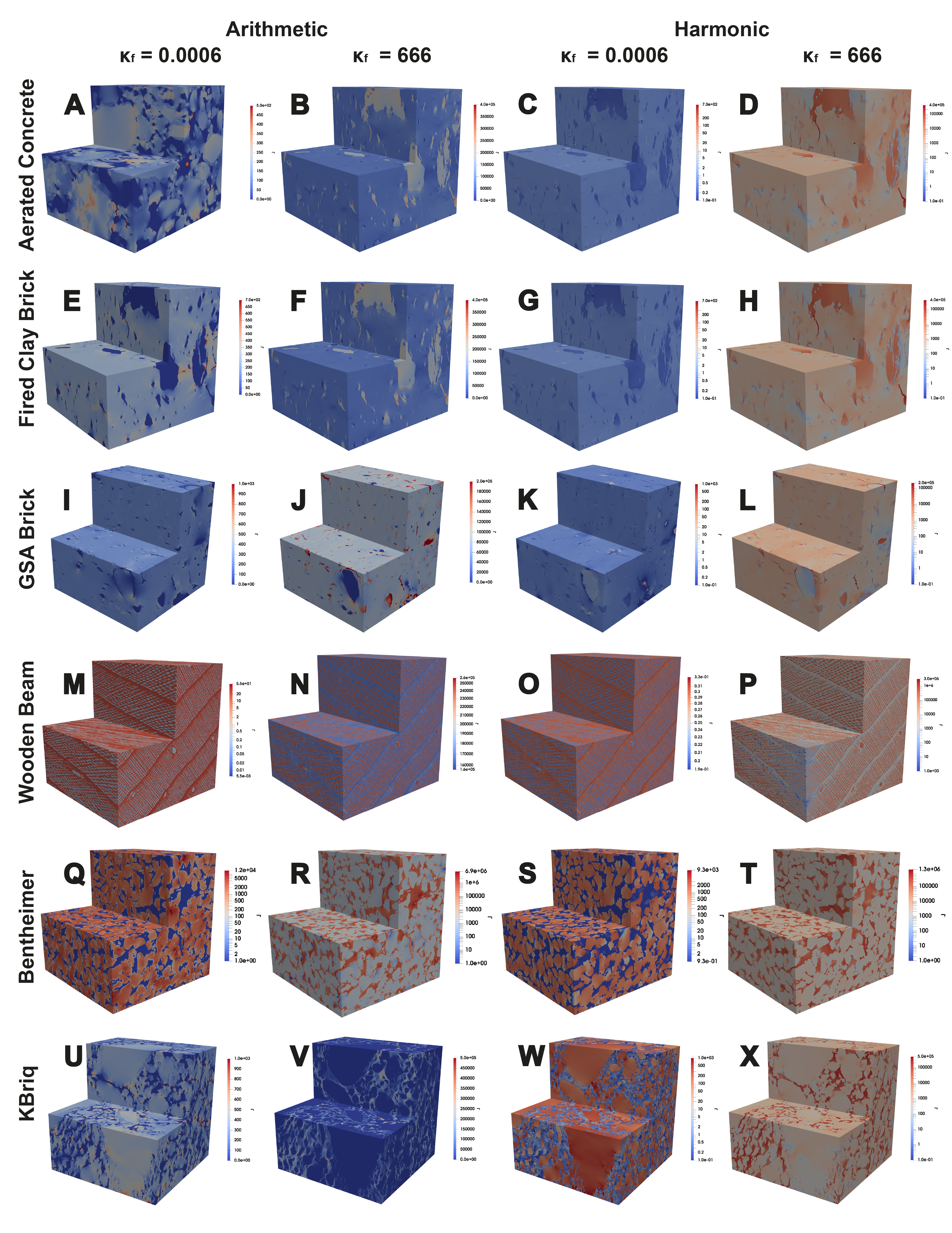}
\caption{ The heat flux J of the simulations for $\mathbf{\kappa}_{f}$ = 0.0006 (columns 1 and 3) and $\mathbf{\kappa}_{f}$ = 666 (columns 2 and 4) using $\mathbf{\kappa}_{\mathbf{\mu}(arithmetic)}$ and $\mathbf{\kappa}_{\mathbf{\mu}(harmonic)}$ for all of the samples. For visual illustration purposes, each simulation is scaled based on linear scaling for arithmetic cases and logarithmic scaling in harmonic cases with the same maximum value where red is high J and blue is low J.  \label{fig:HeatJ}}
\end{center}
\end{figure*}

As the true thermal conductivity of a microporous solid is dependent on a structure that  can not be known in sub resolution region without imaging at another scale, harmonic or arithmetic averaging of that region is a reasonable approximation. However, it has yet to be investigated as to if there is a difference in overall calculated $\mathbf{\kappa}_{eff}$ when each method is used and indeed which method produces more accurate results under which circumstances.  For each sample two solver configurations were then used: (1) $\mathbf{\kappa}_{eff}$ calculated with arithmetic averaging of $\mathbf{\kappa}_{s}$ and $\mathbf{\kappa}_{f}$ for the microporous solid phase and (2) $\mathbf{\kappa}_{eff}$ calculated with harmonic averaging $\mathbf{\kappa}_{s}$ and $\mathbf{\kappa}_{f}$  for the microporous phase.  Input $\mathbf{\kappa}_{s}$ values for the simulations are shown in Table \ref{Table:HeatProperties}.  Heat Transfer was then simulated on the samples for fluids ranging from $\mathbf{\kappa}_{f}$ = 0.0006 to $\mathbf{\kappa}_{f}$ = 666. Fig. \ref{fig:HeatJ} shows the J flux plotted for each of the samples for the highest and lowest values of $\mathbf{\kappa}_{f}$. The results of each simulation shown in Fig. \ref{fig:Keff} are then compared to the theoretical value of $\mathbf{\theta}_{eff}$, corresponding to the theoretical limit as calculated with the same method:

\begin{equation}\label{Eq:thetaharm}
\mathbf{\theta}_{eff(harmonic)} = \frac{\mathbf{\kappa}_{f}\mathbf{\kappa}_{s}}{((1-\mathbf{\phi}_{total})\mathbf{\kappa}_{f}+\mathbf{\kappa}_{s}\mathbf{\phi}_{total})} 
\end{equation}
\begin{equation}\label{Eq:thetaarith}
\mathbf{\theta}_{eff(arithmatic)} = (1-\mathbf{\phi}_{total})\mathbf{\kappa}_{s}+\mathbf{\phi}_{total}\mathbf{\kappa}_{f}
\end{equation}

It is important to note that while in typical systems the pore fluid will be dominated by air or water with a low $\mathbf{\kappa}_{f}$, in extreme weather events such as floods this may not always be the case with penetration of mud or contaminant laden fluids, possibly even metal particulates. Thus we have included a large range of $\mathbf{\kappa}_{f}$ values to span a range of possible pore fluids. 

In Bentheimer (Fig. \ref{fig:Keff}. E ), it is observed that all simulations the  $\mathbf{\kappa}_{eff}$  tracks the $\mathbf{\theta}_{eff(arithmatic)}$. This is because there is no porosity in the solid and thus fluid inside the solid phase to modify the  $\mathbf{\kappa}_{\mathbf{\mu}}$  of the solid. In addition, both grains and pores are well connected so the heat transfer can be modelled  conceptually as a combination of two stacked phases as is analogous to arithmetic averaging. 

For Aerated Concrete good agreement is observed between $\mathbf{\kappa}_{eff}$  and  $\mathbf{\theta}_{eff}$  when  $\mathbf{\kappa}_{f}$ < $\mathbf{\kappa}_{s}$ . However, when $\mathbf{\kappa}_{f}$ > $\mathbf{\kappa}_{s}$ both simulations $\mathbf{\kappa}_{eff}$  track the  $\mathbf{\theta}_{eff(arithmatic)}$ . This is because the fluid pathways are large and well connected with little tortuosity, so when $\mathbf{\kappa}_{f}$ is high, those pathways can be used for heat transport and the porous medium can be effectively modelled arithmetically. However, when  $\mathbf{\kappa}_{f}$ is low, most of the heat is transported through the grains and then the simulations follow the exact law in the solver. Additionally,  using the harmonic function pulls the $\mathbf{\kappa}_{\mathbf{\mu}}$  towards the smallest of  $\mathbf{\kappa}_{f}$ and $\mathbf{\kappa}_{s}$ . The solid inside the microporous phase thus behaves like a barrier. 

For the GSA Brick good agreement is observed between $\mathbf{\theta}_{eff}$  and  $\mathbf{\kappa}_{eff}$  at all  $\mathbf{\kappa}_{f}$  except for a small deviation at high  $\mathbf{\kappa}_{f}$  in the harmonic case when the fractures act as a heat conduit and some pull towards $\mathbf{\theta}_{eff(arithmatic)}$ is seen. However, the fractures are only a small portion of the sample and thus this deviation is only small. The Clay Fired Brick shows very similar behavior where vugs and cracks act as faster heat conduits. However, there are enough of them to impede heat flow in the solid when  $\mathbf{\kappa}_{f}$ < $\mathbf{\kappa}_{s}$ and thussome deviation of  $\mathbf{\kappa}_{eff(harmonic)}$   from  $\mathbf{\theta}_{eff(harmonic)}$ is observed in these cases. 

In the Wooden Beam , $\mathbf{\theta}_{eff}$  and  $\mathbf{\kappa}_{eff}$  are in agreement when  $\mathbf{\kappa}_{f}$ < $\mathbf{\kappa}_{s}$. However, when  $\mathbf{\kappa}_{f}$ is high, the harmonic simulations track  $\mathbf{\theta}_{eff(arithmatic)}$. This is attributed to the long and thin pores that act as direct conduits. Yet, these pores do not stretch completely end to end, so the fluid has to pass through some solid and in the harmonic case this results in much lower $\mathbf{\kappa}_{eff}$ for the solid phase. 

For the KBriq a similar trend to Aerated concrete when $\mathbf{\kappa}_{f}$ > $\mathbf{\kappa}_{s}$ is observed. Nevertheless,  there is a smaller influence of the microprosity when $\mathbf{\kappa}_{f}$ < $\mathbf{\kappa}_{s}$, which is attributed to a lower overall porosity and thus greater ability of the solid phase to transmit heat through the grains.  

It is thus concluded that in many cases knowledge of both the macro and micro structures are required when choosing how to model heat transfer. In cases where there is no microporosity and the solid and grains are well-connected, the arithmetic function should be used. Still, when microporosity is present, the structure of the macropores must be considered. The macro pores themselves can act as either barriers or conduits to heat flow depending on their arrangement and the ratio between  $\mathbf{\kappa}_{f}$ and $\mathbf{\kappa}_{s}$. Furthermore, use of arithmetic or harmonic averaging of the microporous phase can change the resulting $\mathbf{\kappa}_{eff}$ by several orders of magnitude in some cases. It is thus posited that some knowledge of the underlying nanostructure within the microporous phase must be incorporated in order to choose an accurate solver method.  

\begin{table*}[h]
\footnotesize
\centering
\newcommand{\wrap}[1]{\parbox{.05\linewidth}{\vspace{1mm}#1\vspace{1mm}}}
\begin{tabular}{c|c|c|c|c|c|c|c|c}
Sample & $\mathbf{\phi}$  & $\mathbf{\kappa}_{s}$ & $\mathbf{\gamma}_{s}$   & $\mathbf{\rho}_{s}$ & U (arithmetic) & U (harmonic) & t (arithmetic) & t (harmonic)\\[0.005cm]
 & $[-]$ & $kW/m/K$ & $kJ/kg/K$  & $kg/m^3$ & $W/m^2/K$ & $W/m^2/K$ & hr & hr\\[0.05cm]
\hline 
& & \\[-0.2cm]
Aerated Concrete & 0.372 & 1.4 & 0.96 & 2400 & 0.14 & 0.29 & 5.67 & 11.78 \\[0.05 cm]
Fired Clay Brick & 0.365 & 1.0 & 1.05 & 1362 & 0.15 & 0.57 & 3.90 & 14.36 \\[0.05 cm]
GSA Brick & 0.359 & 1.0 &1.05 & 1362 & 0.16 & 0.62 & 3.97 & 15.85 \\[0.05 cm]
Wooden Beam & 0.819 & 0.2 & 2.00 & 600 & 1.18 & 1.44 & 7.16 & 8.75 \\[0.05 cm]
KBriq & 0.179 & 1.4 & 0.96 & 2400 & 0.11 & 0.15 & 5.83 & 7.90 \\[0.05 cm]
Bentheimer & 0.226 & 3.3 & 0.92 & 2650 & 0.07 & 0.08 & 3.54 & 4.17 \\[0.05 cm]
\end{tabular}
\caption{Thermal and porosity properties of the samples used in the heat transfer simulations. As the makeup of a KBriq is proprietary, the same values for density and specific heat capacity as Aerated Concrete were used. The calculated $\mathbf{U_{eff}}$ for air is shown for both arithmetic and harmonic as well as the time for heat to diffuse through a 0.1 m wall.  \label{Table:HeatProperties}}
\end{table*}

\subsection{Upscaling to real systems}

It is then possible to extrapolate these results to a real system by considering a wall of 0.1m thickness $\mathbf{L}$. Here the $\mathrm{U-value}$ of the system $\mathrm{[Wm^{-2}K^{-1}]}$ can be approximated as $\mathbf{L}/\kappa_{eff}$. The time $\mathbf{t}$ for heat to diffuse through the wall is equal to $\mathbf{L^2/D}$ where $\mathbf{L}$ is the wall thickness and $\mathbf{D}$ = $\mathbf{\kappa}_{eff}/ [\rho_{s}(1-\theta)\gamma_s+\rho_{f}\theta\gamma_{f}]$ where $\mathbf{\gamma}$ is the specific heat capacity in $\mathrm{[kJkg^{-1}K^{-1}]}$. A summary table of the calculate times using both the Keff from arithmetic and harmonic averaging and resultant U values and diffusion time are shown in Table \ref{Table:HeatProperties}. Here it is shown that, depending on geometry and the presence of microporosity the difference in calculated U value and heat diffusion time can vary between almost no change in the case of Bentheimer (with no microporosity), to almost a 5-fold change in the cases of the Fired Clay and GSA Bricks, where microporosity dominates. This again illustrates the importance of understanding the microporous structure for effective upscaling of these parameters. 

\section{Conclusions}
A novel workflow for upscaling permeability, flow, and thermal conductivity in a range of porous building materials is presented. Six building material were imaged with micro-CT and analyzed for microporosity and connectivity. The flow and permeability solved on the images both with and without microporosity. Finally, the effective thermal conductivity was solved using both harmonic and arithmetic averaging of the microporous phase and the U value and diffusive time was calculated for a 0.1m thick wall. 

A strong dependence is found on the connectivity of the pore structure for both permeability and thermal conductivity. When the pores were well connected the microporous phase did not effect permeability appreciably. However, in cases where the microporosity connects otherwise disconnected macropores, it must be included to get an accurate measure of permeability and flow. Indeed, in some cases where the macropores are wholly disconnected, permeability cannot be computed without including the microporous phase. 

Thermal conductivity was also effected by local heterogeneity and the arrangement of pores, where macro pores act as either heat conduits or barriers depending on the ratio of thermal conductivity between the microporous solid and liquid phases. When the macropores are not well-connected, the choice of arithmetic or harmonic averaging of thermal conductivity in the sub-resolution porosity can change the effective thermal conductivity by order of magnitude. Furthermore, this in turn can change the calculated U-values and heat diffusion timescale of the bulk materials by up to half an order of magnitude. It is thus imperative that the nano structure is investigated to inform this decision and will be a target for future work. 

This is the first study to combine micro-CT imaging and multi-scale modelling of flow and heat transfer that has applications in the sustainable building industry. These techniques open up the possibility of using this workflow to streamline the design custom materials for optimum permeability and insulation properties for individual use cases. In addition, this workflow could easily be adapted to understand and improve designs in other industries that use porous materials such as fuel cells and batteries technology, lightweight materials and insulation, and semiconductors as well as multi-scale structures from the natural world such as termites nests\cite{singh2019architectural}.

\section{Acknowledgements}
This work was generously funded by the EPSRC project EP/P031307/1. H.P.M. and K.M.H. would also like to thank the Glasgow School of Art for permission to use their samples.


\bibliographystyle{cas-model2-names}

\bibliography{cas-refs}

\end{document}